\documentclass[lettersize,journal]{IEEEtran}
\usepackage{amsmath,amsfonts}
\usepackage{array}
\usepackage[caption=false,font=normalsize,labelfont=sf,textfont=sf]{subfig}
\usepackage{textcomp}
\usepackage{stfloats}
\usepackage{url}
\usepackage{verbatim}
\usepackage{graphicx}

\usepackage{csquotes}
\usepackage{algorithm}
\usepackage{algpseudocode}
\usepackage{amsmath}
\usepackage[most]{tcolorbox}
\usepackage{enumitem}
\usepackage{pifont}
\usepackage{hyperref}
\usepackage{booktabs}
\usepackage{pgfplots}

\definecolor{dkgreen}{rgb}{0,0.6,0}
\definecolor{gray}{rgb}{0.5,0.5,0.5}
\definecolor{mauve}{rgb}{0.58,0,0.82}
\definecolor{mygray}{gray}{0.9}
\colorlet{lightblue}{blue!70}
\colorlet{lightred}{red!50}
\colorlet{redframe}{red!50}
\definecolor{frame}{rgb}{0.5961, 0.7647, 0.9412} 
\definecolor{backg}{rgb}{0.94, 0.97, 0.99}
\definecolor{myred}{rgb}{0.952,0.506,0.506}
\definecolor{mygray}{rgb}{0.512,0.512,0.512}
\definecolor{myblue}{rgb}{0.584,0.882,0.827}

\definecolor{lightred}{HTML}{FFEBEB}

\def\BibTeX{{\rm B\kern-.05em{\sc i\kern-.025em b}\kern-.08em
    T\kern-.1667em\lower.7ex\hbox{E}\kern-.125emX}}
\usepackage{balance}
\begin{document}
\title{A hybrid LLM workflow can help identify user privilege related variables in programs of any size}
\author{
    \IEEEauthorblockN{Haizhou Wang\IEEEauthorrefmark{1}, Zhilong Wang\IEEEauthorrefmark{1}, Peng Liu\IEEEauthorrefmark{1}}\\
    \IEEEauthorblockA{\IEEEauthorrefmark{1}Pennsylvania State University
    \\\{hjw5074, zzw169, pxl20\}@psu.edu}
}


\maketitle

\begin{abstract}
Many programs involves operations and logic manipulating user privileges, which is essential for the security of an organization.
Therefore, one common malicious goal of attackers is to obtain or escalate the privileges~\cite{provos2003preventing, jaafar2016systematic}, causing privilege leakage.
To protect the program and the organization against privilege leakage attacks, it is important to eliminate the vulnerabilities which can be exploited to achieve such attacks.
Unfortunately, while memory vulnerabilities are less challenging to find, logic vulnerabilities are much more imminent, harmful and difficult to identify.
Accordingly, many analysts choose to find user privilege related (UPR) variables first as start points to investigate the code where the UPR variables may be used to see if there exists any vulnerabilities, especially the logic ones.
In this paper, we introduce a large language model (LLM) workflow that can assist analysts in identifying such UPR variables, which is considered to be a very time-consuming task.
Specifically, our tool will audit all the variables in a program and output a UPR score, which is the degree of relationship (closeness) between the variable and user privileges, for each variable.
The proposed approach avoids the drawbacks 
introduced by directly prompting a LLM to find UPR variables by focusing on leverage the LLM at statement level instead of supplying LLM with very long code snippets.
Those variables with high UPR scores are essentially potential UPR variables, which should be manually investigated.
Our experiments show that using a typical UPR score threshold (i.e., UPR score $>$0.8), the false positive rate (FPR) is only 13.49\%, while the number of UPR variables found is significantly more than that of the heuristic based method.

\end{abstract}

\section{Introduction} \label{sec:introduction}
Many organizations and companies run server-side programs to provide services to various users, and therefore, it is necessary to properly control the privileges each user may possess.
Typically, user privileges related operations implemented in a server-side program include but not limited to authentications and authorizations (e.g., the mapping between user IDs and file/data permissions).
Such user privilege related operations are often critical in terms of the security of the programs, the servers, and the service providers. 
Specifically, service providers need to carefully control the privileges granted to users, preventing the privilege leakage risks. 
Thus, in ideal cases, these privilege related operations should be implemented carefully (by very experienced programmers) to guard these operations from being affected by 
memory corruptions (e.g., buffer overflows) and exploitations of logic vulnerabilities. 
However, it is observed in the real world that many privilege related operations are not adequately guarded. 

Since many privilege related operations are not yet adequately guarded, 
many service providers regularly review and analyze their source code to find vulnerabilities that may cause user privilege leakage. 
In general, the relevant vulnerabilities can be categorized into two types: memory corruption vulnerabilities and logic vulnerabilities.
Memory corruptions at runtime, such as buffer overflows and integer overflows, 
are regularly exploited by attackers. 
By providing crafted payload can an attacker modify certain values in 
the program memory, and therefore launch various attacks, including 
control data attacks (e.g. ROP~\cite{rop, rop-no-ret}) and 
non-control data attacks (e.g. DOP~\cite{dop}).
Logic vulnerabilities are vulnerabilities related to the program logic only, and the corresponding exploitations usually do not result in any memory corruption errors. 
One type of logic vulnerability of privilege leakage is 
hard-coded secrets, such as API keys, tokens, etc. 
Some hard-coded secrets can be very inconspicuous.
For example, if a programmer logs the plaintext passwords into the log file, then the passwords can be seen by any person who has access to this log file, essentially causing privilege leakage. 
\autoref{code:logging} illustrates such a case, which we will discuss in detail in \autoref{sec:motivation}.
Another type of logic vulnerability is related to web resource access in which a portion of a URL represents the location of a file/resource on a web server. This vulnerability makes the path traversal attack possible so that an attacker can manipulate the path in a URL to access files or directories that they shouldn't have access to.

Although service providers are strongly motivated to 
identify such logic vulnerabilities, this is not a trivial task.
On the one hand, the program's execution and functionality are not interrupted by such privilege-related logic bugs, which makes it very challenging to detect them through automatic bug finding tools (such as Fuzz testing tools) relying on crashes or sanitizers to detect the abnormal program behaviors. 
On the other hand, since such vulnerabilities are blended in the program logic, 
a good understanding of certain higher-level program semantics is usually required.  
As a result, large tech companies still rely on code review process to discover and eliminate such vulnerabilities.  

In order to make the code review process successful, 
the very first challenge to solve is to spot the privilege-related operations 
and code statements. 
As the privilege operation logic is diverse, real-world analysts 
usually find user privilege related ({\bf UPR}) variables first instead of 
directly spotting the code statements. 
Take the popular static analysis tool \texttt{semgrep} used in industry as an example, more than 40\% of their detection rules focus on specifying the patterns 
that are related to UPR variables. 
Another merit of finding UPR variables instead of the code statements is that 
some variables contain sensitive information, so that the variables themselves need 
to be investigated.
However, finding UPR variables is not an easy task either, since there are still various types of UPR variables, such as passwords, API keys, tokens, etc.

Although it is very important to identify UPR variables, 
existing works unfortunately mainly rely on 
heuristic rules (such as regular expressions), whose 
scalability is limited when screening large code bases or porting to new applications. 
For example, many \texttt{semgrep} rules leverage regular expressions to match the potential keywords that could exist in UPR variables. 
When using regular expressions to discover UPR variables, one 
can easily foresee that this approach is likely to suffer from a high false negative rate.
In addition, based on our experiments in \autoref{sec:evaluation:heuristic}, we also found that this approach could suffer from a high false positive rate as well.
The unsatisfying performance of heuristic-based methods is mainly 
caused by the lacking of explicit patterns for UPR variables.
Specifically, there is no easy way to summarize all the features and 
characteristics of UPR variables, as whether a variable is UPR largely 
depends on higher-level application logic. 
For example, although HTTP servers often have file and CGI paths, 
FTP servers and SSH/Telnet servers don't. 

\noindent\textbf{Key Observation.} Since user privileges are related to 
the logical level of program understanding, it is necessary to analyze 
the application logic of the program before accurately discovering 
UPR variables. 

However, understanding the role of each variable in terms of 
the business/application logic of the program is non-trivial.  
Several previous work~\cite{feng2020codebert, guo2020graphcodebert, li2021palmtree} strive to leverage pre-trained language models to gain understanding of the program given the code; however, the gained understanding is not very satisfying. Fortunately, since the emergence of large language models (LLM) such as ChatGPT, program comprehension tasks become less challenging, and therefore it becomes more feasible to achieve AI-assisted UPR variable identification. 

Despite the promising performance shown by LLMs, it is still non-trivial to build a practical LLM workflow to detect UPR variables. 
As we will shortly show in \autoref{sec:motivation:naive}, even for one of the most powerful LLMs, GPT-4, researchers will encounter challenges when directly prompting a LLM to find UPR variables given the source code. 
Specifically, it is widely observed that as the input context length grows, the performance of the LLM deteriorates~\cite{press2022train, chen2023extending, xiao2023efficient, liu2023lost,jimenez2023swe}, so that identifying UPR variables in long code snippets becomes less effective.  

In this paper, we propose a novel LLM workflow that can help analysts to identify UPR variables in a program with any size. 
Specifically, {\em our approach combines static program analysis and LLM to audit each variable in a program}, so that each variable will be given a UPR score, which can reflect the degree of relevance of a variable to user privileges.
Our approach avoids the drawbacks 
introduced by directly prompting a LLM to find UPR variables by focusing on leverage the LLM at statement level instead of supplying LLM with very long code snippets.  
The UPR score will be scaled between 0 to 10, where 0 means the 
variable is completely not related to user privileges. 
After the UPR scores are given to all the variables, analysts will be able to investigate those variables with higher scores to confirm whether a variable is UPR.
To summarize, our contributions include:
\begin{itemize}
    \item We propose a novel LLM workflow that can help human analysts to identify UPR variables in a program of any size.
    \item We have evaluated the proposed method by investigating the reported potential UPR variables manually. According to our experiments presented in \autoref{sec:evaluation:fp}, when only investigating the top variables with highest UPR scores ($>$8.0), only 13.49\% false positive rate (FPR) is reported, and only less than 15\% of the variables need manual investigation.
With that being said, analysts may save up to 85\% of the time comparing to manually checking all the variables in a program while looking for UPR variables.  
    \item We have compare the proposed method with the widely-adopted heuristic based method, whose heuristic rules are included. It is shown that our method can identify significantly more UPR variables while maintaining much less FPR. 
\end{itemize}

\section{Background} \label{sec:background}
\subsection{Program Dependency Analysis}
Program dependence analysis is fundamental for various program analysis tasks, such as program slicing~\cite{weiser1984program}, and program comprehension~\cite{yu2001hidden}.
It is used to identify the control and data dependence relations between statements (at source code level) or instructions (at binary level) in a program.
Control dependence describes a situation in which whether a statement will be executed is affected by another statement.
If whether statement A will be executed depends on the execution result of statement B, then statement A is said to be control dependent to statement B.
Data dependence describes a relation between statements based on variable definition and usage.
If statement A used a variable defined at statement B, then statement A is data dependent on statement B.

Typically there exists numerous pairs of such dependence relations, and therefore most program dependence analysis methods rely on the program dependence graph~\cite{ferrante1987program} (PDG) to represent both control and data dependence.
Each nodes in the source-code level PDG could be either code statement, expressions, or variable.
Intuitively, nodes that are connected by an edge are considered to have very close relationship, so that program dependence relations can be used to track the "side effect" of a code statement.

\subsection{Large Language Model in Software Security}
Large language models (LLMs) have been attracting attentions from researchers of various fields, after the release of ChatGPT.
Since then, researchers in the field of software security and software engineering have started exploring how LLM may assist solving various tasks, such as program repair~\cite{kang2023large,jimenez2023swe}, program testing~\cite{lemieux2023codamosa}, and fuzzing~\cite{xia2023universal}.
As more and more researchers start working on this field, the contributions of most of works can be categorized into 3 categories: 1) prompt engineering, 2) LLM fine-tuning, and 3) workflow design.

Though LLMs are promising in helping solving software challegnes, adopting LLM could be non-trivial. 
For example, despite the diversity of papers using LLM to solve software challenges, many of them involve using LLM to generate code.
Indeed LLM is very powerful in generating code, but it is also true that the quality of generated code snippets are usually unpredictable~\cite{jimenez2023swe}.
Furthermore, the quality of the LLM output can be greatly affected by the context length~\cite{press2022train, chen2023extending, xiao2023efficient,liu2023lost,jimenez2023swe}, which introduces challenges when working with long code snippets.

\section{Motivations and Problem Statement} \label{sec:motivation}
\subsection{Importance of UPR Variables}


\definecolor{dkgreen}{rgb}{0,0.6,0}
\definecolor{gray}{rgb}{0.5,0.5,0.5}
\definecolor{mauve}{rgb}{0.58,0,0.82}
\definecolor{mygray}{gray}{0.9}

\lstset{
language=C,
frame=tb,
basicstyle={\footnotesize \bf \ttfamily}, 
tabsize=3,
breaklines=true,
showstringspaces=false,
numbers=left,
numbersep=-10pt,                     
escapeinside={(*}{*)},
xleftmargin=2pt,
numberstyle=\texttt \footnotesize,
stringstyle=\color{mauve},
keywordstyle= \color{black}, 
commentstyle=\color{dkgreen} \textit,
}

Many server side programs involves data operations, such as CRUD operations in web servers.
Since a great number of users may have the access to the server, it is necessary to limit the privileges each user have, so that they are only allowed to access and/or modify the data for which they have authorization.
The simplest example is the credential information of user accounts.
If the the password of a user is leaked, an attacker can easily get all the privileges that user bear.

Accordingly, goals of many cyberattacks are essentially gaining or escalating privileges (i.e. privilege leakage), and therefore it is crucial to patch vulnerabilities which can be exploit to achieve privilege leakage.
Since both memory~\cite{rop, ret2libc, jop} and logical vulnerabilities~\cite{data-attack, dop} may be exploit to achieve potential privilege leakage, finding such vulnerability in the program becomes non-trivial.
In particular, while some existing methods such as fuzzing may be used to find memory vulnerabilities automatically, there is no effective ways to find logic bugs.
In many cases, just inappropriate coding practices may expose the privilege related information in danger, such as hard-coded tokens in source code managed by version control systems (e.g., git).
\begin{figure}[h] 
  \begin{lstlisting}[language=go,
      emph={},
      caption={\textbf{An example of improper logging practice.} },label=code:logging,captionpos=b]
        // improper logging practice
        logging.INFO("%s", password)
        ...
  \end{lstlisting}
\end{figure}
\begin{figure}[h] 
  \begin{lstlisting}[language=python,
      emph={},
      caption={\textbf{An example of hardcoded secrets.} },label=code:hardcode,captionpos=b]
        ...
        # hardcoded secrets
        api_key = "YOUR_OPENAI_API_KEY"
        response_round_1 = openai.Completion.create(
            engine=model,
            prompt=prompt_round_1,
            max_tokens=150
        )
        ...
  \end{lstlisting}
\end{figure}
\autoref{code:logging} and \autoref{code:hardcode} illustrates two simple cases where logic vulnerability may lead to privilege leak.
In \autoref{code:logging}, the program will log the plaintext password into the log file, which may be accessed by other users.
If the leaked password is obtained by an attacker, obviously certain privileges will be available to the attacker.
In \autoref{code:hardcode}, an API key is hardcoded in the source code, which could be catastrophic if the source code is under version control. 
In the events of the version control synchronizing with an remote server, many users will be able to access this hardcoded API key.

Therefore, to efficiently review the privileges handling logic in a program, real-world analysts usually start from identifying those variables containing information related to user privilege.
However, it is impractical to manually analyze every single variables in the programs, especially for organizations who have huge code bases and many proprietary components.

\subsection{Existing Methods to Identify UPR Variables} \label{sec:motivation:challenges}
Only without considering the cost of time is it possible to have analysts manually identify all the UPR variables.
For example, a medium size program ($<$10K lines of code) may contain thousands of variables, so that analyzing them one by one could be extremely time consuming and exhausting.

On the other hand, automatically identifying UPR variables is very challenging, because they are closely related to the high-level semantics of the program.
In other words, whether they are UPR is partially, if not completely, depending on the functionality of the program.
Therefore, no explicit general pattern can be found for the UPR variables.

Consequently, existing UPR variable detection methods are mostly heuristic-based, and a typical strategy adopted by real-world companies is to implement such heuristics using commercial static analysis tool, such as \texttt{Semgrep}, to handle their large-scale code bases.
For example, a very common heuristic is to detect UPR variables by investigating the names of UPR variables, because UPR variables may contains certain user privilege related keywords (e.g. key, token, password, etc.).
According, this variable name matching heuristic will be often implemented as regular expression.

Unfortunately, such heuristic-based methods can only cover cases which are specified in the heuristic rules.
Using the variable name matching heuristic as an example, this strategy obviously will omit those UPR variables with less informative names.
In addition, a variable containing user privilege related keywords is not necessarily a UPR variables.
For example, usually a variable named \texttt{token} could be a UPR variable, but a variable named \texttt{token\_type} is not likely to be a UPR variable.
Furthermore, creating such heuristics requires human analysts to enumerate as many possibilities as possible, causing scalability issues.

Another challenge is that the influence of a UPR variable could be propagated to other variables, making other variables to be UPR.
This implies that the dependencies between variables are also important when determine whether a variable is UPR.
Thus, there also exist taint-analysis based heuristics to detect UPR variable, which set user privilege related API calls (e.g., \texttt{setuid}, \texttt{setgid}, etc.) as taint sinks.
However, such detection method can only be used to find UPR variables involving the UPR API calls, leaving a substantial number of UPR variables undetected.

As a result, from analysts' perspective, it is desirable to have an automatic UPR variable detection tool that can save them time and efforts, while reporting as many UPR variables as possible.
Not only the name of the variables, but also the dependence relationships in the program should an ideal UPR variable detection tool consider.

\subsection{Using LLM Naively} \label{sec:motivation:naive}
As elaborated in \autoref{sec:motivation:challenges}, in order to identify UPR variables, it is important to take the contexts of inter-variable dependencies and program functionalities into consideration.
As LLMs start showing their strength in understanding the natural language and program code, they seem to be a potential solution to identify UPR variables automatically.

The very natural idea of finding UPR variables using LLM would be simple: just ask it.
We have conducted preliminary experiments of \textit{prompting the LLM directly to find UPR variables}.
Specifically, the experiments are conducted using GPT-4, which is considered to be the one of the most powerful LLM.
To provide the GPT-4 with all the contexts, the prompt starts with an initial prompt providing the instructions and specification of the UPR variable, followed by the related source code without any modifications.
In the experiments, we used \texttt{bftpd}, an medium size open-source FTP server, as the test program.
We designed our initial prompt, which is shown in \autoref{fig:naive-prompt}.
\begin{figure}
    \centering
    \begin{tcolorbox}[
  colback=backg,colframe=frame,colbacktitle=red!80!black ]
I want you to act as a security critical variable identifier. I will type a code snippet and you will tell me which variables are critical variables. Regarding what is a critical variable, when it is tampered, it may result in:
\begin{enumerate}
    \item[1.] One can bypass authentication and/or authorization
    \item[2.] One can achieve behaviors that are being banned
    \item[3.] One can gain elevated privileges (uid changes)
    \item[4.] One can access files/data that are not accessible (chown related)
    \item[5.] One can steal secrets (keys, passwd, etc.)
\end{enumerate}
\end{tcolorbox}
    \caption{Initial prompt to find UPR variables using naive method}
    \label{fig:naive-prompt}
\end{figure}

We use the notion of "critical variable" along with a list of specification criteria, in an effort to clarify what is a UPR variable.
For an FTP server, the listed 5 criteria should cover most of the UPR variables.
In this experiment, we decided to conduct analysis at \textit{function level}.
We have manually tested roughly 25 functions, and the experiments had mixed results.
In particular, we observe for shorter functions (i.e., less than 50 lines) involving only a few variables, the results looks promising, so that all the UPR variables are identified with only very few false positives; however, when analyzing longer functions, the result is not satisfactory. For example, when we ask GPT-4 to analyze the function \href{https://github.com/gamman/bftpd/blob/cc1df5805918f65f1f419cd79b228dc045f34e51/login.c#L190}{\texttt{bftpd\_login}} in \texttt{login.c} for program \texttt{bftpd}, it gives unsatisfying reply, which is shown in \autoref{fig:naive-prompt-reply}.

\begin{figure}
    \centering
    \begin{tcolorbox}[colback=lightred,colframe=redframe]
        The critical variables in this code snippet are:\\
        1. \texttt{password}: This is the password provided by the user for authentication. If tampered, it can lead to unauthorized access.\\
        2. \texttt{file\_auth}: This variable points to the file used to authenticate users. If tampered, it can lead to unauthorized access or bypassing of authentication.\\
        3. \texttt{home\_directory}: This variable holds the home directory retrieved from the auth file. If tampered, it can lead to unauthorized access to files/data.\\
        4. \texttt{anonymous}: This variable holds the configuration option for anonymous user. If tampered, it can lead to unauthorized access or bypassing of authentication.\\
        5. \texttt{change\_uid\_text}: This variable holds the configuration option for changing user id. If tampered, it can lead to privilege escalation.\\
        6. \textcolor{red}{\texttt{user}}: This variable holds the username of the user trying to login. If tampered, it can lead to unauthorized access or bypassing of authentication.\\
        7. \texttt{userinfo.pw\_uid} and \texttt{userinfo.pw\_gid}: These variables hold the user id and group id of the user. If tampered, it can lead to privilege escalation.\\
        8. \texttt{str}: This variable is used to store various critical information like root directory, user name, etc. If tampered, it can lead to unauthorized access to files/data.\\
        9. \textcolor{red}{\texttt{config\_getoption("DO\_CHROOT")}}: This configuration option decides whether to change the root directory or not. If tampered, it can lead to unauthorized access to files/data.\\
    \end{tcolorbox}
    \caption{GPT reply when prompted naively}
    \label{fig:naive-prompt-reply}
\end{figure}

In this response, in total 4 UPR variables are missing and 2 identified are false positives, including:
\begin{enumerate}
    \item \texttt{user} is not UPR variable, marked in red.
    \item \texttt{config\_getoption("DO\_CHROOT")} is not even a variable, marked in red.
    \item \texttt{state} is missing, which is used to indicate the user is authenticated in this function.
    \item \texttt{change\_uid} and \texttt{change\_uid\_text} are missing, which is a configuration that can affect the \texttt{home\_directory} and the execution of \texttt{set\_uid} call.
    \item \texttt{anon\_ok} is missing, which is a configuration to allow anonymous users.
\end{enumerate}

Consequently, we have derived a very important observation: \textit{as the code snippet grows longer, the capability of the LLM in detecting UPR variables will decrease}.
Unfortunately, GPT-4 is completely a blackbox, so that its parameters and meta-information are not available.
Nevertheless, our observation does echo with many previous works.
According to the experiments in previous works~\cite{press2022train, chen2023extending, xiao2023efficient} focusing open-sourced LLMs, their results show that as the input length approach the context window size of an LLM, the perplexity of the output will increase.
In cases of normal chat, the perplexity can be viewed as reversed-fluency, which may be less of an issue for the purpose of communication.
However, identifying UPR variable is a reasoning task, so that such "surprise" is not acceptable.
Another work~\cite{liu2023lost} conclude that the LLMs are \textbf{not} robust to changes in the position of relevant information: specifically that LLM will fail to retrieve information if the relevant information is in the middle of the input context.
As the function length grows, the potential UPR variables and its relevant code statements could "lost in the middle".
In the field of LLM assisted program analysis, similar observation could be found.
In SWE-Bench~\cite{jimenez2023swe}, authors also observe that when using LLMs to resolve GitHub issues (i.e., code repair task), Claude 2 and other LLM's performance drops as the input context length increases.

\subsection{Problem Statement}

Despite the mixed results shown in \autoref{sec:motivation:naive}, it is clear that LLMs can help with circumventing the challenge of lack of explicit pattern for UPR variables.
However, as shown in \autoref{sec:motivation:naive}, the LLM is neither panacea nor silver bullets, so that adopting it naively could lead to failure in real world settings.

We aim to develop a UPR variable detection method that can significantly reduce the amount of time needed to identify UPR variables. Specifically, to address the challenges and issues mentioned in \autoref{sec:motivation:challenges}, our problem statement is as followed:
\textit{Given a program and a capable LLM, how to leverage the LLM to find the UPR variables in the program, so that substantially more UPR variables could be identified while maintaining a reasonable false positive rate.}

Based on our experiments and observations, the most imminent challenge when adopting LLM to identify UPR variables is:
when finding UPR variables in large code bases, the long input context length will deteriorate the robustness of LLM, leading to false positives and false negatives.

A straightforward solution would be trimming the input function body by using, for example, information retrieval~\cite{robertson2004simple, jimenez2023swe}.
But it is difficult to decide what and how much code to be removed so that enough information is left to determine whether a variable is UPR.
Furthermore, even if there exists a perfect trimming strategy, the remaining context of a variable could be already too long for an LLM to perform robustly.

In this paper, we propose a noval LLM-assisted workflow, which can output the \textbf{UPR score} of a variable, to assist a human analyst in identifying UPR variables.
The UPR score is a quantitative metric assessing the degree of relationship between a variable and user privileges.
To alleviate LLMs' performance deterioration issue related to the input context length, we only leverage LLMs to generate the \textbf{UPR ratings for code statements} instead of whole chunks of code, so that the input context length of the prompts could be minimized.
Other necessary aspects to determine the UPR score of a variable are addressed by traditional program analysis techniques, specifically by constructing the program dependence graph (PDG)~\cite{ferrante1987program, austin1992dynamic} at code statement level.

\section{Method} \label{sec:method}

\begin{figure*}
    \centering
    \includegraphics[width=\linewidth]{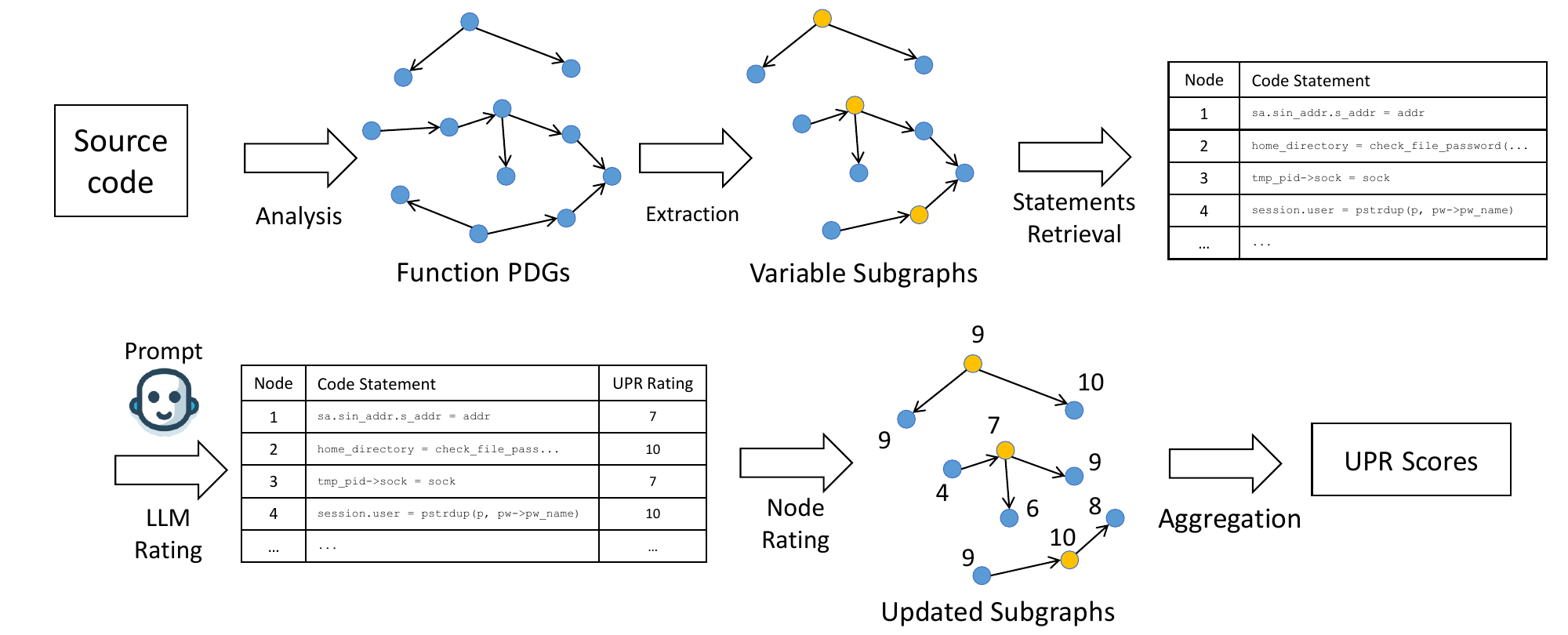}
    \caption{Workflow}
    \label{fig:workflow}
\end{figure*}

\subsection{Issue of Binary Verdict} \label{sec:method:bin}
Before introducing the detail of our workflow, we would like first elaborate the reasons why we choose to output a UPR score instead of a UPR verdict.
Essentially, the diversity among the UPR variables are substantial, which implies that they may have different degrees of relationship to the user privileges.
Therefore, some variables in a program may be at the borderline, meaning that they could be either UPR or not, depending on the context of the program/function as well as the stored values at runtime.

\begin{figure}[t] 
  \begin{lstlisting}[language=C,
      emph={file_uid},
      caption={\textbf{An example of a borderline UPR variable.} },label=code:borderline_example,captionpos=b]
        int get_file_uid(char* filename) {
            int file_uid; // Is this UPR?
            ...
            return file_uid;
        }
        int do_ls(char* cmd) {
            ...
            int uid = get_file_uid(fname);
            ...
        }
        int do_chown(char* cmd) {
            ...
            int uid = get_file_uid(fname);
            if (uid == current_uid) {
                ...
                chown(pathname, new_uid, new_gid);
                ...
            }      
            ...
        }
  \end{lstlisting}
\end{figure}

\autoref{code:borderline_example} shows an example of such cases.
In an FTP server, a common utility function is to read the \texttt{stat} information of files, including uid and gid.
Unlike passwords, uids are less sensitive and just leaking the values of the uids may not cause severe consequences.
Typically, a variable saving uids is UPR only when it is used as the argument of function calls such as \texttt{chown} or \texttt{setuid}.
Let us assume a function called \texttt{get\_file\_uid}, whose returned value is going to be stored in a local variable called \texttt{file\_uid}.
This function could be used by both the \texttt{ls} command to list the dir for the user, and the \texttt{chown} command to change the owner for a file.
Needless to say, only in case of \texttt{chown} command will the variable \texttt{file\_uid} be UPR.
In this case, it is very difficult to make a UPR verdict for the local variable \texttt{file\_uid} in function \texttt{get\_file\_uid}.

Although the borderline cases are relatively rare, they can distract the analysts if not being properly handled.
By outputting a score rather than a verdict would alleviate this issue, and fortunately, the powerful LLMs make this possible.

\subsection{Workflow Overview} \label{sec:method:overview}
\autoref{fig:workflow} illustrates the complete workflow of our approach, which include 6 major steps.
Starting from the source code of a target program as input, the workflow eventually output the UPR scores for all variables in the program.

\textcircled{1} First, from the source code files, we generate the PDG for each function through static analysis. 
Each nodes in the PDG corresponds to a statement expression. 
\textcircled{2} The PDGs are then sliced into variable subgraphs, each of which corresponds to a single variable in the program. 
\textcircled{3} Next, based on the variable subgraphs, we collect the code statements, creating a set of code statements that are either control and data dependent on or by the variables. 
\textcircled{4} This set of code statements are subsequently rated by an LLM, using an initial prompt that is designed upon the purpose of the target program.
The ratings shall reflect the significance of the statements in terms of their impacts toward potential non-control data attacks against the target program. 
\textcircled{5} Then, the statement ratings will be used to update the variable subgraphs, serving as node score.
\textcircled{6} Finally, we compute the score of the variable by aggregating the node scores using a aggregation algorithm based on the graph topology. To confirm a UPR variable, subsequent manual review is necessary for those with highest ratings.

\subsection{PDG Generation} \label{sec:method:pdg}
Since source code level PDGs are very well defined~\cite{ferrante1987program}, we adopt the well-known definition of PDG and will not reiterate any formal definitions and specifications.

Essentially, we parse the source code files and generate a PDG for each function. 
As discussed in \autoref{sec:motivation}, to control the length of the input length of the prompts as well as to balance the information carried by the graph topology and nodes, our PDG is designed to only have \textbf{simple statements} as nodes, including \texttt{assignment}, (function) \texttt{call}, \texttt{assertion}, \texttt{goto}, and \texttt{return}. 
In cases of a compound statement (i.e.,block of simple statements), including \texttt{for}, \texttt{while}, \texttt{do} loops, \texttt{switch}, \texttt{if}, \texttt{else} branches, etc., the condition is treated as a single node (i.e., simple statement), and each simple statement in the block will also be a single node, with an control dependence edge connected to the condition node.

\begin{figure}[t]
    \centering
    \includegraphics[width=\linewidth]{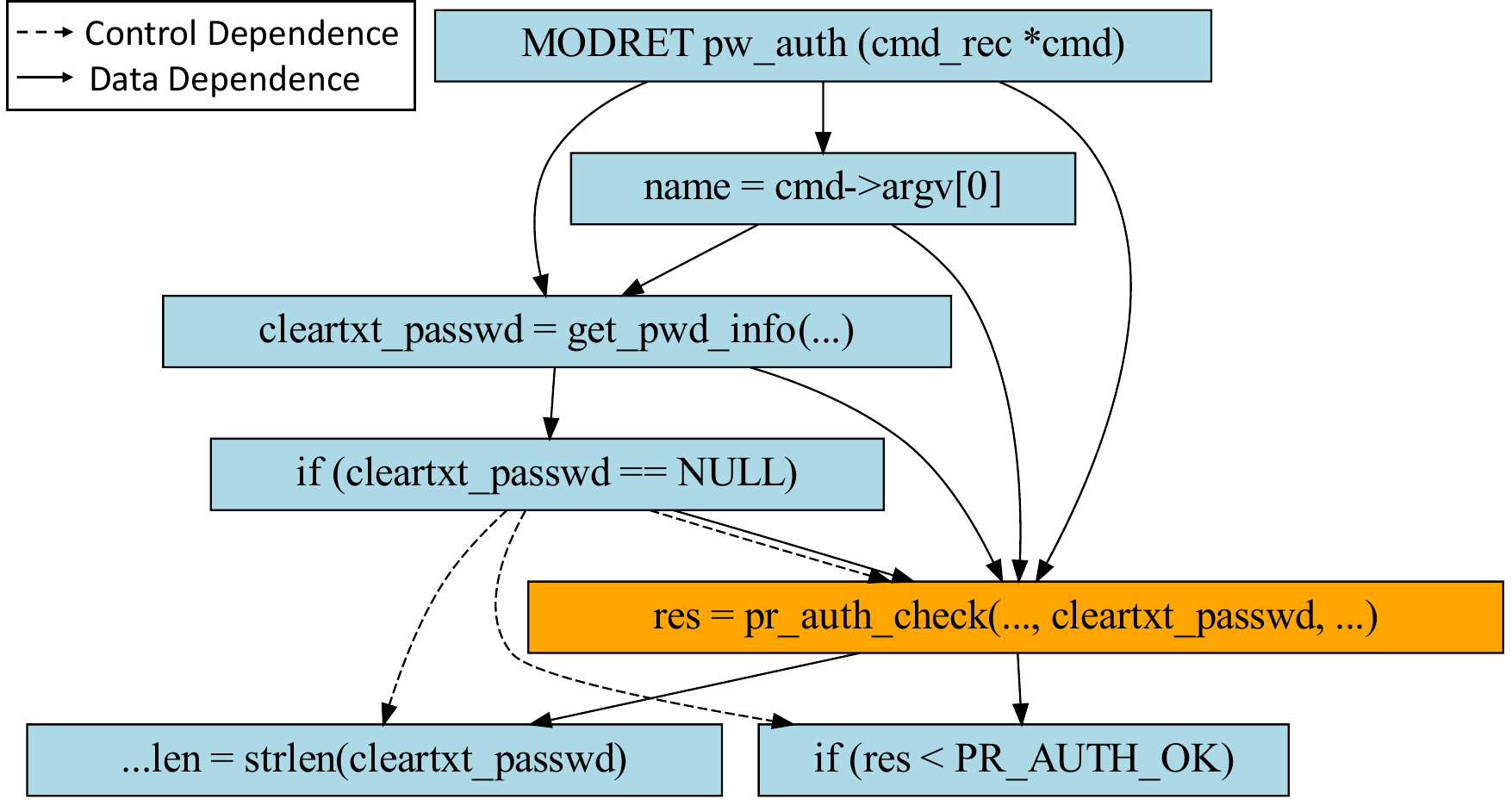}
    \caption{A pruned variable subgraph for a single criterion node of variable \texttt{cleartxt\_passwd} in program \texttt{proftpd}.}
    \label{fig:ddg_example}
\end{figure}

\subsection{Variable Subgraph Extraction} \label{sec:method:vargraph}
From the function PDGs, we then extract variable subgraphs for each referred variable in the function, including global variables and local variables. 
Operating on the PDG, the extraction process essentially can be viewed as a customized program slicing~\cite{weiser1984program} process, so that we slice the PDG in both backward and forward directions from the criterion nodes.

First, we identify all the variables referred in the function through AST analysis.
For each variable, we then mark all the nodes corresponding to the statements referring it as criterion node (for program slicing).
Since a variable could be referred multiple times in the function, intuitively each variable could have multiple criterion nodes, so that between a variable and its criterion nodes is there a one-to-many relation, as shown in \autoref{fig:workflow}.
Next, we extract the variable subgraph for each variable by exclusively including only immediate (i.e., 1-hop) neighbors of the corresponding criterion nodes, regardless the direction.
One one hand, both directions are imperative, as whether a variable is UPR depends not only on its depender statements, but also on the dependee statements;
on the other hand, statements that are not immediate depender/dependee of the criterion nodes are trivial, if not detrimental, to determine whether a variable is UPR, as further dependency relations are more noisy and less relevant.
\autoref{fig:ddg_example} shows an pruned example of the extracted variable subgraphs, which only contains a single criterion node.
Note that nodes are represented in different colors, where orange ones are criteria nodes.
The variable subgraphs are extracted from the PDG, and therefore they have two types of edges: control and data dependence edges.

Since nodes in the PDG are statements, it is completely possible that two variables are referred by a same set of statements, rendering identical variable subgraphs for different variables. 
In case one of them were UPR but the other were not, either false positive or false negative will present.
This issue is observed in our experiments, and we have it discussed in \autoref{sec:evaluation:fp}

\subsection{Code Statement UPR Rating Using LLM} \label{sec:method:rating}
After variable subgraphs are constructed, we collect code statements from all the subgraphs, creating a set of statements to be rated by the LLM. 
To clarify, since the statements are collected from the variable subgraphs, there is no sequential order between the statements in the set. 
The LLM will evaluate and rate the statements one after another, so that the ratings of the statements are independent to each other. 
Each statement in the set is unique, so that the cost of using the LLM can be minimized.

To instruct the LLM with necessary information and context, our prompts consist two parts: an initial prompt consist of instructions, and the code statement to be rated. 
The initial prompt serves for two goals: 1) prompt the LLM to generate ratings for each statements; and 2) provide the specifications of the UPR property.
The detail of the prompt design will be discussed in \autoref{sec:prompt}.

Indeed, LLM is built upon neural networks after all, and thus it is a black-box.
Therefore, the ultimate UPR ratings for code statement given by the LLM may be based on not only the criteria in the initial prompt, but also LLM's own "interpretations".
With that being said, the code statement UPR ratings given by the LLM could be disagreed by a human analyst, so that our workflow need to be designed to tolerate this inconsistency.

After the code statements are rated, effectively their corresponding nodes in the variable subgraphs will receive UPR ratings, which can be further used to determine the UPR scores of the variables.

\subsection{Generating Variable UPR Score}
The output of our workflow is a UPR score for each variable, which is obtained by aggregating the node scores in the updated variable subgraph.
Recall that our variable subgraph is essentially a pruned PDG, so that the topology of the graph implies the dependence relationships between the code statements.
As discussed in \autoref{sec:motivation:challenges}, whether a variable is UPR pivots on not only the how the variable is used and defined, but also how the variable will affect other part of the program.
Thus, the UPR score is computed by considering all the nodes in the updated variable subgraph.

We formalize the procedure of generating UPR scores for variables:
Given a updated variable subgraph $G=(N,E)$ of variable $v$, where $N$ is the set of nodes and $E$ is the set of edges in the graph, and a parameter $\lambda$ weighing between the node score and its neighbors' score, we can compute the UPR score of a variable, as shown in \autoref{algo:upr_score}.

\begin{algorithm}
    \caption{Compute UPR score}
    \label{algo:upr_score}
    \begin{algorithmic}[1]
        \Require{$G=(N,E)$, $\lambda \in [0,1]$}
        \State $S=\{\}$ \Comment{Array to store aggregated score of all criteria nodes}
        \For{$n \in N$}
            \If{$n$ is criterion node}
                \State $R=\{\}$ \Comment{Array to store score of neighbors}
                \State $s_c = \textrm{get\_node\_score}(n)$
                \State $V = \textrm{find\_neighbor\_nodes}(n)$
                \For{$v \in V$}
                    \State $R = \{R, \textrm{get\_node\_score}(v)\}$ \Comment{Append}
                \EndFor
                \State $s_v = \textrm{max}(R)$
                \State $s_n = \textrm{Normalize}(s_c + \lambda s_n)$ \Comment{Aggregated score}
                \State $S = \{S, s_n\}$ \Comment{Append}
            \EndIf
        \EndFor
        \State \Return $\textrm{max}(S)$
    \end{algorithmic}
\end{algorithm}

\autoref{algo:upr_score} is essentially a aggregation process, whose most important operation is at line 10, 11 and 15.
There are two aggregating operations.
The first one is at line 11 of \autoref{algo:upr_score}, where we compute the aggregated score of each criterion node.
Do note that this score of a criterion node is affected by both the criterion node itself and its neighbors. 
Here, the value of $\lambda$ is between 0 and 1, and the larger the more important neighbor score is.
The normalize function ensures final UPR score is in a fix range, that is, between 0 to 10.
The second aggregating operation is at line 15, so that we obtain a single UPR score for a variable from scores of multiple criteria nodes.
It is worthwhile to note that, as mentioned in \autoref{sec:method:rating}, our workflow needs to tolerate potential inconsistent ratings with human.
Based on the principle of reduce false negatives, we use \texttt{max} as our aggregation function.

After collecting UPR scores for all variables, the variables will be sorted by their UPR scores from high to low.
An analyst can then set up a threshold so that any variables with a UPR score above the threshold will considered as UPR variable candidates, subject to further manual analysis.
As such, analysts do not have to manually check every single variables in a program.

\section{Prompting LLM} \label{sec:prompt}
A fundamental component of our workflow is an LLM, which helps rate a code statement in terms of
the degree of the relationship between the code statement and user privileges.
Obviously, the quality of the rating result is one of the most important factors toward generating appropriate UPR scores and therefore helping analysts to identify UPR variables.
However, there is no principled method to design prompts, due to the black-box nature of LLM.
Therefore, we designed our prompt based on empirical case studies: we have selected 30 code statements from different programs of different lengths, and then gradually improve the prompt based on the human feedback of how well the UPR ratings are for the selected code statements.

As briefly introduced in \autoref{sec:method:rating}, the prompt consists of two parts: initial prompt and the code statement to be rated.
Since the code statement has to be presented as it is, the prompt engineering effort is mainly for the initial prompt.
Eventually, we decide to adopt the following template for initial prompt:
\begin{tcolorbox}[
  colback=backg,colframe=frame,colbacktitle=red!80!black ]
I want you to act as a security critical code statement identifier. I will type a code statement and you will reply with a criticalness rating from 0 to 10 of the code statement. 0 is the least level of criticalness. Only reply with the criticalness rating and do not write explanation. Regarding what is a critical code statement, when it is tampered, it may result in:
\begin{enumerate}
    \item[1.] **Malicious Goal 1**
    \item[2.] **Malicious Goal 2**
    \item[3.] ...
\end{enumerate}
\end{tcolorbox}

In this design, after describing our request to the LLM, we include a list of potential malicious goals that an attacker may be interested related to the user privileges.
The list of malicious goals can be compiled based on the target programs, as different programs may have different kinds of user privilege related issues.
In this paper, we compiled a list of common user privilege security incidents, generating a list of malicious goals used in the prompt:
\begin{tcolorbox}[
  colback=backg,colframe=frame,colbacktitle=red!80!black ]
\begin{enumerate}[leftmargin=10pt]
    \item[1.] One can bypass authentication and/or authorization
    \item[2.] One can achieve behaviors that are being banned
    \item[3.] One can gain elevated privileges (uid changes)
    \item[4.] One can access files/data that are not accessible (chown related)
    \item[5.] One can steal secrets (keys, passwd, etc.)
\end{enumerate}
\end{tcolorbox}

As will discussed shortly in \autoref{sec:evaluation}, the UPR security incidents in selected programs in our experiments can be mostly covered by this list.
In addition, the effect of the prompt design will be also studied and discussed in \autoref{sec:evaluation:prompt}

\section{Experiments and Evaluation} \label{sec:evaluation}
We evaluate our approach in terms of both practicality and reliability, proposing 4 research questions: 1) How practical our approach is in terms of the false positive rate?
2) Can our approach find more UPR variables than heuristic-based method?
3) How to interpret distribution of the rating spectrum?
4) What is the effect of the prompt design?
5) How does it perform on proprietary programs? These research questions will be answered in later sub-sections.

Our evaluation uses a code base constructed using in total 7 programs: \texttt{bftpd}, \texttt{proftpd}, \texttt{vsftpd}, \texttt{ghttpd}, \texttt{nginx}, \texttt{sshd}, \texttt{sudo}.
The programs are selected to ensure diversity in both program functionality (i.e., business logic) and size (i.e., lines of code). 

To reduce both time cost and economical cost of our experiments, we did not analyze all the variables in the selected program.
Specifically, for programs with a small code base, we run all variables though our method; whereas for larger programs, we only include variables in the core modules (so that a minimum functional binary can be compiled).
For example, \texttt{sshd} is one of the binary in the \texttt{OpenSSH} code base, so that in our experiments, all the variable in modules for other binaries (e.g., \texttt{ssh}, \texttt{ssh-keygen}, \texttt{scp}, etc.) are excluded.

\subsection{Experiment Setup and Implementation}
The prototype of our tool leverages several existing tools. From the source code, the raw PDG is extracted using \texttt{Joern}.
The extracted raw PDG is further processed using Python scripts to generate variable subgraphs.
Subsequently, we use GPT-4 as the LLM code statement rater, as it is one of the best LLMs available.
We send each statement along with the initial prompt discussed in \autoref{sec:prompt} to the GPT-4, and collect the UPR ratings evaluated by the LLM.
Finally, the UPR score of the variable is computed based on \autoref{algo:upr_score} implemented in Python.
 
To reduce the number of variables may be missed, we conduct static analysis to identify variables in the program using both LLVM (at LLVM IR level) and \texttt{Joern} (at AST level).
In total, the prototype of our tool used in the experiments are implemented in more than 1600 lines of code in Python and 1300 lines of code in C/C++.

\subsection{How practical is our approach regarding FPR?} \label{sec:evaluation:fp}
It would be meaningless if the proposed method could not reduce manual efforts of the analyst.
In particular, we want the analyst to find as many UPR variables as possible by examining as less variables as possible.
Thus, without an adequate false positive rate (FPR) will never our approach be practical.

We evaluate the practicality of our approach in terms of the FPR by experiments.
Simulating how our proposed tool may be used in real world, 
we first take the target program and evaluate the variables using our workflow discussed in \autoref{sec:method:overview}.
After obtaining the UPR scores of each variable, we set up a rating threshold of 9.0 (recall that the output UPR scores are between 0 and 10), above which are considered as UPR variable candidates.
Then, the UPR variables are going through a manual analysis: only those variables which contain sensitive information, or/and can be exploited to cause security incidents (e.g., bypass authentication, gain additional privileges, etc.)
are marked as UPR.

\begin{table}[ht]
    \centering
    \begin{tabular}{cccc}
    \toprule
    Program    & Analyzed & Detected & FPR    \\
    \midrule
    bftpd   & 551      & 60        & 0.1500 \\
    proftpd & 1153      & 235       & 0.1069 \\
    vsftpd  & 829      & 73       & 0.1311 \\
    ghttpd  & 71       & 3        & 0.0000 \\
    nginx   & 708        & 15        & 0.1333 \\
    sshd    & 679      & 195       & 0.1485 \\
    sudo    & 464       & 64        & 0.1905 \\
    \midrule
    All     & 4455     & 645      & 0.1349 \\
    \bottomrule
    \end{tabular}
    \vspace{2mm}
    \caption{False Positive Rate of Our Method}
    \label{tab:fpr_all}
\end{table}

The experimental results are shown in \autoref{tab:fpr_all}.
Since the selected programs are diverse regarding both their functionalities and their size, it is not surprising that the number of UPR variables varies between programs.
We only conduct false positive analysis because manually identifying UPR variable is extremely time-consuming.
By only focus on false positives can we only inspect 645 \textit{Detected} variables.
Were we to compute the false negative rate, instead 4455 \textit{Analyzed} variables need to be inspected.

The results suggest that the FPRs are 13.49\% on average.
In other words, for every 100 of variables reported, less than 14 of them will be false positive.
Nevertheless, when we take a glance at the number of variable \textit{analyzed} in \autoref{tab:fpr_all}, without our tool, in total 4455 of variables will need to be manually checked comparing only 645 are now need to be checked.
Thus, we believe our tool can significantly reduce the manual efforts required from the human analysts.

After careful analysis of the false positives, we found that the most common cause for false positives is due to the granularity of the PDG.
As mentioned in \autoref{sec:method:vargraph}, it is totally possible that two variables have almost or exactly identical variable subgraphs.
Using an example shown in \autoref{fig:ddg_example}, \texttt{cleartxt\_passwd} and \texttt{cleartxt\_passwd\_len} (abbreviated in \autoref{fig:ddg_example} to save space).
In code statements where \texttt{cleartxt\_passwd\_len} is used, many of them also will use \texttt{cleartxt\_passwd}.
Eventually, after the variable subgraph extraction, the variable subgraph for \texttt{cleartxt\_passwd} and \texttt{cleartxt\_passwd\_len} is similar.
Unfortunately, though the password itself is clearly a UPR variable, the length of it is not, and therefore the similar variable subgraphs will cause the false alarm.

\subsection{Can our approach find more UPR variables than heuristic-based method?} \label{sec:evaluation:heuristic}

One key motivation of our method is to identify more UPR variables than heuristic-based methods.
To answer this research question, we design an experiment to compare our method with heuristic-based methods commonly adopted in the industry.
Specifically, heuristic-based methods usually adopt two kinds of heuristics: 1) code pattern matching and 2) taint analysis. 

Accordingly, we design a set of heuristic rules that can be applied using \texttt{Semgrep} framework, which is an open-source static analysis tool to find bugs and security vulnerabilities in the source code.
It allows developers to write custom rules using a simple syntax to search for patterns in codebases.
Since it supports syntax matching and taint analysis, we develop two rules containing several heuristics to discover UPR in the source code, as shown in \autoref{code:syntax} and \autoref{code:taint}.

    \begin{lstlisting}[
        language={[x86masm]Assembler},
        emph={},
        morekeywords={mode,pattern-source,inside,either,patterns,pattern,pattern-inside, metavariable-regex,metavariable,regex,pattern-sinks,pattern-either},
        deletekeywords={if},
        caption={Syntax based detection rules for UPR.},
        keywordstyle=\color{blue}\bfseries,
        label=code:syntax,captionpos=b]
    patterns:
      - pattern-either: 
          - pattern-inside: |
              if (...) 
          - pattern-inside: |
              switch (...) 
      - pattern: $VARIABLE
      - metavariable-regex:
          metavariable: $VARIABLE
          regex: .*(?<![A-Za-z])(auth|authenticate|login|admin|authorize|banned|allowed|uid|gid|username|permission|chown|key|passwd|password)(?![A-RT-Za-rt-z]).*
    \end{lstlisting}

    \begin{lstlisting}[
        language={[x86masm]Assembler},
        emph={},
        morekeywords={mode,source,inside,either,sinks,patterns,pattern,pattern-inside, metavariable-regex,metavariable,regex,pattern-sinks,pattern-either},
        keywordstyle=\color{blue}\bfseries,
        caption={Taint analysis based detection rules for UPR.},
        label=code:taint,captionpos=b]
    mode: taint
    pattern-source:
      - patterns:
          - pattern-either: 
              - pattern-inside: |
                  $IOFUNC(...)
              - pattern-inside: |
                  $VAR = $IOFUNC(...)
          - metavariable-regex:
              metavariable: $IOFUNC
              @\Hilightc@regex: scanf|gets|fscanf|fgetc|fgets|getchar|fread
    pattern-sinks:
      - patterns:
          - pattern-inside: |
              seteuid (...) 
          - pattern-inside: |
              setuid (...)
    ...
    \end{lstlisting}

Both rules are straightforward, so we only provide brief explanations.
Specifically, the rule shown in \autoref{code:syntax} detects the variables that contain privilege-related keywords. The most important item in this rule is the regular expression at line~10 matching privilege-related keywords in the variable names:
\begin{lstlisting}[
    language={[x86masm]Assembler},
    emph={},
    numbers=none,
    frame = none]
    .*(?<![A-Za-z])(auth|authenticate|login|admin|authorize|banned|allowed|uid|gid|username|permission|chown|key|passwd|password)(?![A-RT-Za-rt-z]).*
\end{lstlisting}

The rule in \autoref{code:taint} is essentially taint analysis, which specifies the taint source and taint sinks. It can detect variables whose taint comes from inputs that are used in user privilege related function calls.

We then perform the UPR variable detection using the heuristic-based method on exactly the same code base described at the beginning of this section.
To make this experiment practical, we only focus on positive variables that is either reported by our method or the heuristic-based method.
Those variables reported by neither methods will be excluded from our analysis.
In order to make a fair comparison, we divide the detected positive variables into the following three sets: (\textbf{Set A}) variables only reported by heuristic-based method, (\textbf{Set B}) variables only reported by our method, and (\textbf{Set C}) variables reported by both methods.
\begin{figure}[t]
    \centering
    \includegraphics[width=0.73\linewidth]{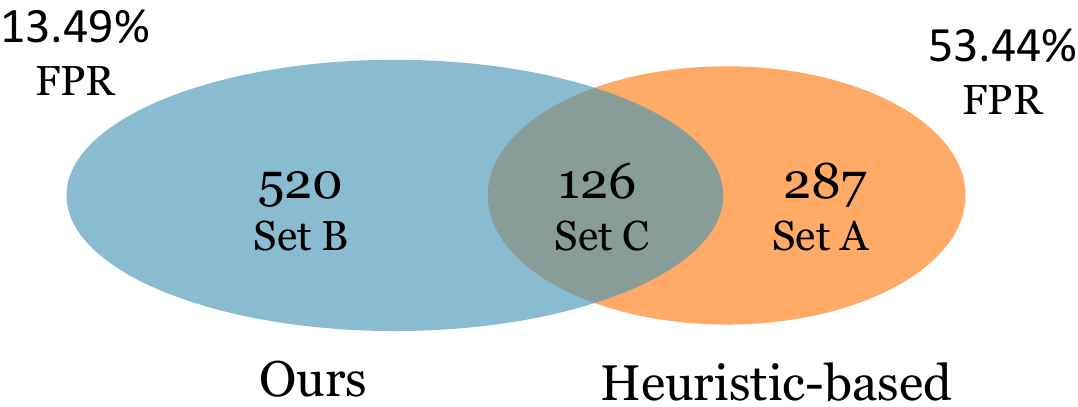}
    \caption{UPR positive variable reported by our method and the heuristic-based method.}
    \label{fig:compare}
\end{figure}
The result of analysis is shown in \autoref{fig:compare}.
According to the overlapped circles in \autoref{fig:compare}, it is clear that our method can find substantially more variables than the heuristic-based method can, especially with a significantly lower false positive rate (13.49\% vs. 54.53\%). 
Do note that our method detected 645 variables as positive, using a fairly conservative threshold (i.e., 9.0/10.0), out of which 13.49\% of them are false positives.
In contrast, when we use the heuristic-based method introduced above, out of 413 (126 + 287), more than half of them (53.44\%) are false positives.
Indeed the heuristic-based method could be tweaked to be more conservative and our method can be more aggressive, but the differences in performance is apparent.

This experiment also exposes some false negative cases for our approach.
A common false negative reason is the quality of the UPR ratings for code statements from GPT-4, specifically the mediocre UPR ratings.
On one hand, the mediocre ratings may be caused by limited capability of the GPT-4; on the other hand, they may be also due to the insufficient amount of the context.
For example, considering following code statement:
\begin{displayquote}
\texttt{if(!access\_check\_file(\&s\_filter\_str))}
\end{displayquote}
Similar to the situation described in \autoref{sec:method:bin}, this is a case of "borderline UPR statement": whether it is UPR cannot be asserted without knowing the value may be stored in \texttt{s\_filter\_str}.
In our experiments, we only consider a variable as a UPR candidate if the UPR score is at least 9, which is a relatively high threshold.
Therefore, a few variables with mediocre-rating code statement in their variable subgraphs are missed.
If such mediocre ratings are very common, it would indeed made our method less practical. We will have more discussions about the UPR ratings from the GPT-4 in \autoref{sec:evaluation:ratings}.

\subsection{How well can an LLM rate the code statement?} \label{sec:evaluation:ratings}
Toward the core of our approach is the UPR rating of the code statements generated by LLMs, which lead to a research question: how well can an LLM rate a code statement in terms of their relationship with user privileges?
Due to the blackbox nature of the LLM, it is futile to inspect the LLM internally.
Since there is no benchmark regarding the task of rating a code statement for LLMs, instead of studying the question directly, we decide to focus on the distribution of the ratings given by the GPT-4 in our experiments.

We firstly set up a series of hypotheses to validate, based on the observations of the UPR variables:
\begin{itemize}
    \item[\textbf{O1}:] UPR variables, especially those ones critically related to user privileges, are rare.
    \item[\textbf{O2}:] The majority of variables are \textbf{not} UPR.
    \item[\textbf{O3}:] A few variables may or may not be UPR, depending on the context of the program. (See \autoref{sec:method:bin})
\end{itemize}
Based on the above-mentioned observations of UPR variables, regarding the UPR scores of code statements, we have following hypotheses:
\begin{itemize}
    \item[\textbf{H1}:] Code statements with higher UPR scores are the minority.
    \item[\textbf{H2}:] The majority of code statements should receive low scores.
    \item[\textbf{H3}:] A few code statements should have mediocre UPR scores, because a single code statement contains no context about the containing function, as well as the caller(s) of the function.
\end{itemize}

Obviously, \textbf{H1} is derived from \textbf{O1}, and it is natural to believe that the number of UPR variables and the number of high-UPR-score code statements are closely correlated.
After all, whether a variable is UPR is depending on whether it is directly or indirectly related to a code statement related to user privileges (i.e. with a high UPR score).
For similar reasons, \textbf{H2} is derived from \textbf{O2}.

\textbf{H3} is derived from \textbf{O3}, which reflect those variables whose UPR scores are depending how they may be used in the program, as well as the business logic of the program.
As discussed in \autoref{sec:method:bin}, though they are uncommon, when analysts encounter these variables, it would be difficult to make a verdict regarding whether the variables are UPR variables.
and the \texttt{chown} command to change the owner for a file.
In our workflow, only the code statement itself is given to the LLM when generating the UPR ratings, so that the context of how the variable is defined and used may be missing.
We therefore expect the LLM may give mediocre ratings in these cases, as a human analyst would also prefer to do so.

\begin{figure}
\begin{tikzpicture}[scale=0.65]
        \begin{axis}[
            width=0.7\textwidth,height=65mm,
            x tick label style={/pgf/number format/1000 sep=},
            ylabel={Count},
            xlabel= {UPR Ratings},
            legend style={
                at={(0.8,0.9)},
                anchor=north west,
            },
            xticklabels={,,0,1,2,3,4,5,6,7,8,9,10},
            nodes near coords,
            every node near coord/.append style={font=\tiny},
            nodes near coords align={vertical},
            ybar=0pt,
            bar width=20pt,
            bar shift=0cm
        ]
        \addplot[color=myblue, fill=myblue]
            coordinates {(0,1964)[{red}] (01,6450) (02,3970)};
        \addplot[color=mygray, fill=mygray]
              coordinates { (03,477) (04,3) (05,534) (06,5) (07,1987)};
          \addplot[color=myred, fill=myred]
            coordinates {(08,2832) (09,426) (10,732)};
        \end{axis}
        
  \end{tikzpicture}
  \label{fig:gpt4ratings}
  \caption{Distribution of the statements UPR ratings from GPT-4}
\end{figure}

\autoref{fig:gpt4ratings} shows the distribution of the UPR ratings given by GPT-4.
The area marked blue are code statements with low UPR ratings, whereas the area marked in red are those ones with high UPR ratings.
The grey areas are considered as mediocre ratings.
According to \autoref{fig:gpt4ratings}, it is evident that all three hypotheses are validated:
\begin{enumerate}
    \item Only few statements have high UPR ratings (small red area).
    \item Large number of statements have low UPR ratings (large blue area).
    \item Only a few statements recieve mediocre ratings (tiny grey area).
\end{enumerate}
Analysis based on the distribution of the UPR ratings may neglect those corner cases.
Nevertheless, \autoref{fig:gpt4ratings} shows that the selected LLM, GPT-4, is capable of generating appropriate UPR ratings for code statements.

\subsection{What is the effect of the prompt design?} \label{sec:evaluation:prompt}
As discussed in \autoref{sec:prompt}, our prompt template include a list of malicious goals to provide the LLM with more specifications about UPR variables.
While the presence of this list can improve the UPR ratings for the code statements, it is intriguing to study whether the presence of this list would affect the \textbf{final UPR scores for variables.}

Accordingly, we evaluate our method using a modified prompt template for code statement UPR ratings acquisition. 
Specifically, the prompt template are modified in two ways: one kind only contains partial list of the malicious goal; whereas the other kind contains no list of malicious goals.
In the modified prompt template containing partial list, two items in the original list are kept: 1) One can access files/data that are not accessible (chown related); 2) One can steal secrets (keys, passwd, etc.)
We use the two modified prompts in our workflow to re-detect the confirmed UPR variables found in previous experiment.

The experiment is evaluated by measuring the percentage of the confirmed UPR variables are re-detected using different modified prompts, which is shown below:
\begin{itemize}
    \item Prompt with partial list: 43.89\%
    \item Prompt with no list: 54.30\%
\end{itemize}
It may seem to be unexpected that the result when using completely no list is better.
As mentioned in \autoref{sec:method:rating}, LLM may have their own "interpretations" for some words or phrases in the prompts.
Therefore, if there is no given specification for UPR variable, the UPR ratings may become very unpredictable. 
For example, in the initial prompt, we used the term "security critical", which is a rather broad phrase, so that GPT-4 can have various interpretation of such a term.
Nevertheless, it is still evident that providing specification such as a list of malicious goal definitely enables our method to uncover more UPR variables.

\section{Discussion and Limitation} \label{sec:discussion}
\subsection{Proprietary Programs}
All the LLMs are trained on massive amount of data, which probably include almost all the open source source code. 
Consequently, from machine learning perspective, essentially the UPR rating for code statements in open source programs are in-distribution data (as oppose to out-of-distribution data).
Though it is believed that LLMs perform well in term of the generalizability, nevertheless out-of-distribution data still may affect the performance of the model.
In our case, the out-of-distribution data is essentially the code statement in proprietary programs.

Since it is not realistic to comprehensively evaluate our approach on proprietary programs, we have conducted a simple experiment about using GPT-4 to re-rate the code statements, but with the names in the statements changed.
For example, a function in \texttt{vsftpd} named \texttt{vsf\_sysutil\_read} will be renamed as \texttt{xyz\_sysutil\_read}, pretending it is a function in different program.
We have randomly selected 30 code statements to have the GPT-4 to re-rate, after renaming all the names.
Our result shows that the average difference between the ratings is -0.45 with a standard deviation of 0.94, which implies that the UPR rating does change, but only moderately.
Nevertheless, thought it is difficult to draw rigorous conclusion regarding the performance on proprietary program, as LLMs are considered as black boxes, this may becomes a limitation when adopting our method to some proprietary programs.

\subsection{Cost of LLM Service}
Our method will prompt LLM for each code statements, so that the number of token used in the input is dominated by the length initial prompt.
While the disadvantage in terms of the cost is that every single statement will need to pay the price of the initial prompt, the advantage is that the output token used, which is typically more expensive than that of input, could be just 1 or 2, because we only want the LLM to output the UPR ratings.

\section{Conclusion}
In this paper, we introduce a large language model (LLM) workflow that can assist analysts in identifying such UPR variables, which is a very time-consuming task.
Our tool will analyze all the variables in a program to output a UPR score for each variable, which is the degree of relationship between the variable and user privileges.
When our approach is used in real world, those variables with high UPR scores will then be manually investigated by the analysts.
Our experiments show that in a typical setting (i.e., UPR score $>$0.8), the false positive rate (FPR) is only 13.49\%, while the number of UPR variables found is significantly more than that of the heuristic based method.

\bibliographystyle{IEEEtran}
\bibliography{main_ieee}

\end{document}